# A New Decomposition Ensemble Approach for Tourism Demand Forecasting: Evidence from Major Source Countries

Chengyuan Zhang[a], Fuxin Jiang[b, c], Shouyang Wang[b], Shaolong Sun [d, *]

[a]School of Economics and Management, Beihang University, Beijing 100191, China;
[b]Academy of Mathematics and Systems Science, Chinese Academy of Sciences, Beijing 100190, China;
[c]School of Economics and Management, University of Chinese Academy of Sciences, Beijing 100190, China;
[d]School of Management, Xi'an Jiaotong University, Xi'an, 710049, China

**Abstract:** The Asian-pacific region is the major international tourism demand market in the world, and its tourism demand is deeply affected by various factors. Previous studies have shown that different market factors influence the tourism market demand at different timescales. Accordingly, the decomposition ensemble learning approach is proposed to analyze the impact of different market factors on market demand, and the potential advantages of the proposed method on forecasting tourism demand in the Asia-pacific region are further explored. This study carefully explores the multi-scale relationship between tourist destinations and the major source countries, by decomposing the corresponding monthly tourist arrivals with noise-assisted multivariate empirical mode decomposition. With the China and Malaysia as case studies, their respective empirical results show that decomposition ensemble approach significantly better than the benchmarks which include statistical model, machine learning and deep learning model, in terms of the level forecasting accuracy and directional forecasting accuracy.

[*] Corresponding author. School of Management, Xi'an Jiaotong University, Xi'an, 710049, China. Email: sunshl@amss.ac.cn (S. L. Sun).

# 1. Introduction

The Asian-pacific region has already developed into the world's major international tourism demand markets and has taken a leading position both arrivals and receipts, owing to a strong economic development, a stable political environment and the technological advances (Chu, 1998; Leung et al., 2011; UNWTO, 2019). Particularly, in this region, the international tourist arrivals and receipts both increased by 7% and recorded the highest growth in arrivals in 2018, reflecting the solid intraregional demand (Leung et al., 2011; UNWTO, 2019). With such growth comes more challenge in ensuring effective destination management that maximize the allocation of tourism resources (Song & Li, 2008; Pan & Yang, 2017). Therefore, to investigate the forecasting performance of rapidly changing tourist arrivals in the Asia-pacific region can not only improve the level of resource management, but also bring enlightenment to the forecast of demand changes in other regions.

Tourism demand has long been extensively affected at different timescales by a wide variety of factors including economic, financial, political environment and other factors and may cause dramatic fluctuations in international tourist arrivals (Chen et al., 2012; Li et al., 2016; Chan et al., 2005; Wang, 2009). For example, Gössling et al. (2012) investigated the short-term effects of climate and weather conditions on tourist trip, especially as tourists may change their bookings or destinations at the last-minute. Day et al. (2013) explored the possible medium-term impact of weather changes on destination demand. Pizam & Fleischer (2002) revealed the severity of the terrorism events and their frequency which have a long-term effect on tourism market. In consequence, demand modelling takes into account the influence of tourism-related factors on tourism fluctuations at different time scales, which will help improve forecast accuracy and promote efficient resource allocation.

Notably, the influencing factors related to tourism demand in major source countries are an important part influencing tourist arrivals at tourist destinations (Tiwari et al., 2018). The analysis of the factors' effect on demand forecasting between tourist destinations and major source countries has been applied by researchers in the past. For example, the policy of visa-free entry for main source countries can produce the positive significance on inbound tourism (Lee et al., 2010; Balli et al., 2013), and a political conflict between two countries significantly damages the country image and increases the negative influence on tourism demand (Alvarez & Campo, 2014). More importantly, it was noticed that foreign tourist arrivals from different countries are dependent on each other (Tiwari et al., 2018; Chan et al., 2005). Therefore, the modelling and analysis of volatility in tourist arrivals in destination's major tourism inbound markets at different timescales can provide a useful tool for improving the forecasting accuracy.

To model the multiscale relationship among the tourist destination and major source market, an emerging decomposition algorithm, i.e., noise-assisted multivariate empirical mode decomposition (NA-MEMD) method, is introduced in this study. Compared with other multi-scale analyses (such as Fourier analysis and wavelet decomposition), NA-MEMD holds the following one unique merit for tourism analysis. Particularly, different from traditional data decomposition techniques using fixed bases, NA-MEMD employs empirical, adaptive bases according to the nature of the studied samples, which is especially suitable for non-stationary, nonlinear and complex data. In view of the above virtue, this study, in particular, introduces this powerful decomposition method, i.e., NA-MEMD, to explore the multi-scale relationship between the tourist destination and the corresponding major tourism inbound markets.

Generally, this study innovatively investigated whether the decomposition among the tourist destination and the corresponding major tourism inbound markets could improve performance of tourism demand forecasting. There are four main steps involve in the proposed methodology: data process, data decomposition, data forecasting and results analysis. First, basic data process is employed for dealing the original data time series. Second, the promising decomposition method, i.e., NA-EMD, is introduced to capture the multiscale relationship between the tourist destination and the corresponding major tourism inbound markets, in terms of extracting matched modes on similar timescales. Third, a popular forecasting technique, either a statistical model or AI technique, is used to conduct individual forecasting on each timescale, then the final forecasting results are obtained in a linear form of different timescales. Fourth, the comparison between the evaluation criteria, the statistical analyses and the robustness analysis are conducted. Comparison to existing studies, this study makes major contributions from the following two perspectives:

(1) This study may be the first attempt to introduce NA-MEMD to capture the multiscale relationship between the tourist destination in Asian-pacific region and the corresponding major

tourism inbound markets, and propose a novel NA-MEMD-based method for tourist arrivals forecasting; and

(2) The effectiveness of the proposed method is empirically verified in comparison with other benchmarks in the tourist volume forecasting of China and Malaysia, respectively.

The rest of this study is organized as follows. The literature review about the data decomposition methods in tourism research and tourist volume forecasting are presented in **Section 2**. **Section 3** describes the proposed methodology, together with the general framework. **Sections 4-5** demonstrate the detail of the experimental design and the empirical study, respectively. **Section 6** concludes the main findings and managerial applications in tourism research.

## 2. Literature review

This section briefly reviews the relevant literature about data decomposition methods in tourism research and the popular forecasting techniques for tourist volume forecasting.

### 2.1 Data decomposition methods in tourism research

In recent years, many researchers have been exploring the different market impact factors drive the volatility of tourism market at different time frequencies (Goh, 2012; Pizam & Fleischer, 2002; Wang, 2009; Wu & Wu, 2018). Accordingly, a number of authors have paid more attention to the application of data decomposition method for analyzing tourism demand from the perspectives of time frequencies, such as spectral analysis (Coshall, 2000; Kožić, 2014), singular spectrum analysis (SSA) (Beneki et al., 2012; Hassani et al., 2015), Fourier decomposition (Apergis et al., 2017), wavelet decomposition (Kummong & Supratid, 2016) and the empirical mode decomposition (EMD) family (Chen et al., 2012; Li & Law, 2019; Li et al., 2016; Zhang et al., 2017a). **Table 1** summarizes the related literature that incorporated data decomposition methods to analyze and forecast tourism demand.

**Table 1**. The literature that incorporated data decomposition method in tourism research.

| References | Region focused | Target | Data frequency | Methods | Variables |
|---|---|---|---|---|---|
| Apergis et al. (2017) | Croatia | Tourism demand | Monthly | Fourier analysis | Tourist arrivals |
| Beneki et al. (2012) | UK | Tourism income | Monthly | SSA | Tourism receipts |
| Coshall (2000) | United States | Inbound demand | Quarterly | Spectral analysis | Exchange rates; Air and sea passengers |
| Chen et al. (2012) | Taiwan | Tourism demand | Monthly | EMD | Tourist arrivals |
| Hassani et al. 2015) | United States | Tourism demand | Monthly | SSA | Tourist arrivals |
| Kožić (2014) | Global | Tourism demand | Annual | Spectral analysis | Tourist arrivals |
| Kummong & Supratid (2016) | Thailand | Tourism demand | Monthly | Wavelet analysis | Tourist arrivals |
| Li et al. (2016) | Mainland China | Tourism demand | Daily | EMD | Tourist arrivals; Baidu Index |
| Li & Law (2019) | Hong Kong | Inbound demand | Monthly | EEMD | Tourist arrivals; Google Trends |
| Lin et al. (2018) | Mainland China | Tourism demand | Daily | EMD | Tourist arrivals |
| Raza et al. (2017) | United States | Tourism demand | Monthly | Wavelet analysis | Tourist arrivals; $CO_2$ emission |
| Wu & Wu (2019) | Portugal, Ireland, Italy, Greece, Spain | Tourism receipts | Annual | Wavelet analysis | European economic policy; Tourism receipts |
| Yahya et al. (2017) | Malaysia | Inbound demand | Monthly | EMD | Tourist arrivals |
| Zhang et al. (2017) | - | Hotel demand | Daily | EEMD | Hotel occupancy rate |
| Zhang et al. (2018) | Charleston, United States | Hotel demand | Weekly | EEMD | Hotel occupancy rate |

SSA: Singular spectrum analysis; EMD: empirical mode decomposition; EEMD: ensemble empirical mode decomposition

Notably, compared with the traditional data decomposition method (i.e., spectral analysis, SSA, Fourier decomposition and wavelet decomposition), the EMD family, which relax the strong assumption (e.g., linearity and stationarity) in the fixed bases used in traditional method, have also been appeared in field of tourism demand forecasting (Li & Law, 2019; Chen et al., 2012; Tang et al., 2020). For example, Chen et al. (2012) proposed a hybrid model with EMD and back-propagation neural network (BPNN) to forecast tourist arrivals and revealed the superiority of decomposition method over the corresponding benchmarks. Yahya et al. (2017) employed the artificial intelligence technique for forecasting each component after decomposition and finally obtained the aggregated forecasting results. Their findings showed that the proposed model can improve the forecasting accuracy. Moreover, as a typical extension of EMD, i.e., the EEMD, has also been introduced in tourist arrival and hotel demand for improving the forecasting accuracy (Li & Law, 2019; Zhang et al., 2017; Zhang et al., 2018). Li & Law (2019) explored the utility of decomposition methods (i.e., EEMD) in extracting components from Google trends data of tourism and verified the effectiveness of introducing the EEMD-based method to improve the tourism demand forecasting accuracy. In the field of hotel occupancy forecasting, Zhang et al. (2017) combined the EEMD method with an autoregressive integrated moving average (ARIMA) for forecasting the daily hotel occupancy, and the hybrid model significantly outperforms the respective benchmarks, especially for short-term forecasts. Similar findings were also obtained by Zhang et al. (2018), who used the modified EEMD-ARIMA-based mothed by grouping the decomposition components for forecasting the weekly hotel demand and achieved better performance.

## 2.2 Tourist volume forecasting

As for the tourist volume forecasting, the popular methods can be roughly divided into three categories involving time series models, econometric models and artificial intelligence techniques, which have been extensively employed in tourism demand forecasting (Song & Li, 2008; Li et al., 2018; Song et al., 2019).

The time series models mainly include the Naïve, Autoregressive moving average (ARMA) and Generalised autoregressive conditional heteroskedastic (GARCH) model, focusing on the corresponding historical patterns (Peng et al., 2014; Song & Li, 2008). Particularly, the autoregressive moving average (ARIMA) as a typical time series method has been widely applied to tourism demand forecasting (Sun et al., 2019; Cho, 2003; Goh & Law, 2002; Jungmittag, 2016; Pan & Yang, 2017; Burger et al., 2001; Chen & Wang, 2007). Moreover, the famous version of ARIMA considering the seasonality (i.e., SARIMA) has also been introduced in the field of tourism demand forecasting and produced fairly accurate forecasts (Chu, 2009; Kulendran & Wong, 2005). Other time series techniques have also appeared frequently for studying the different issues in tourism forecasting studies, such as volatility of tourism demand (Chan, Lim & McAleer, 2005), seasonal pattern in per person tourist spending (Koc & Altinay, 2007) and short-term forecasts on tourist arrivals (Gounopoulos et al., 2012).

To investigate the causality relationship between tourism demand and the influencing factors, the econometric methods have widely emerged as the main forecasting methods in the field of tourism demand forecasting. For example, the error correction model (ECM), the vector autoregressive (VAR) model, the almost ideal demand system (AIDS) and the autoregressive distributed lag model (ADLM) could be found in Wong et al. (2007), Li et al. (2004), Assaf et al. (2019) and Song et al. (2011). Meanwhile, the different versions of the common econometric models are often used in tourism demand forecasting, such as ECM-AIDS (Li et al. (2006), Bayesian VAR (Wong et al., 2006) and AR-MIDAS (Bangwayo-Skeete & Skeete, 2015).

In recent years, artificial intelligence (AI) techniques as a powerful tool for modelling the non-linear data have been applied popularly to forecast non-stationary, nonlinear and complex time series, especially in the case of tourism demand analysis (Sun et al., 2019; Song et al., 2019). Particularly, the fuzzy logic theory, artificial neural networks (ANNs), support vector regression (SVR), genetic algorithms (GA), back-propagation neural network (BPNN), extreme learning machine (ELM) and Long Short-Term Memory (LSTM) have emerged frequently in tourism analysis field (Wang, 2004; Burger et al., 2001; Chen & Wang, 2007; Sun et al., 2019; Law et al., 2019). For example, Chen & Wang (2007) employed the support vector regression (SVR) method for forecasting the quarterly tourism time series in China and performed well. Li et al. (2018) introduced search engine data as input in BPNN for improving the forecasting performance. Furthermore, the LSTM as a famous extension of recursive neural network has also been applied in tourism demand forecasting (Law et al., 2019).

The review of the literature shows data decomposition methods have been used to forecast tourist arrival to reduce the complexity and improve forecasting accuracy. However, none of the existing studies have employed the multivariate data decomposition method considering the major source countries for effectively exploring the multiscale relationship among the countries with strong ties. Therefore, this study attempts to fill in such a literature gap to present a novel decomposition ensemble approach for improving the accuracy of tourism demand forecasting.

## 3. Methodology

### 3.1 General framework

In this section, we propose a novel decomposition ensemble approach (i.e., NA-MEMD-based method) for tourism demand forecasting, as shown in **Fig. 1**. The framework describes the modelling process with data process, data decomposition, data forecasting and results analysis.

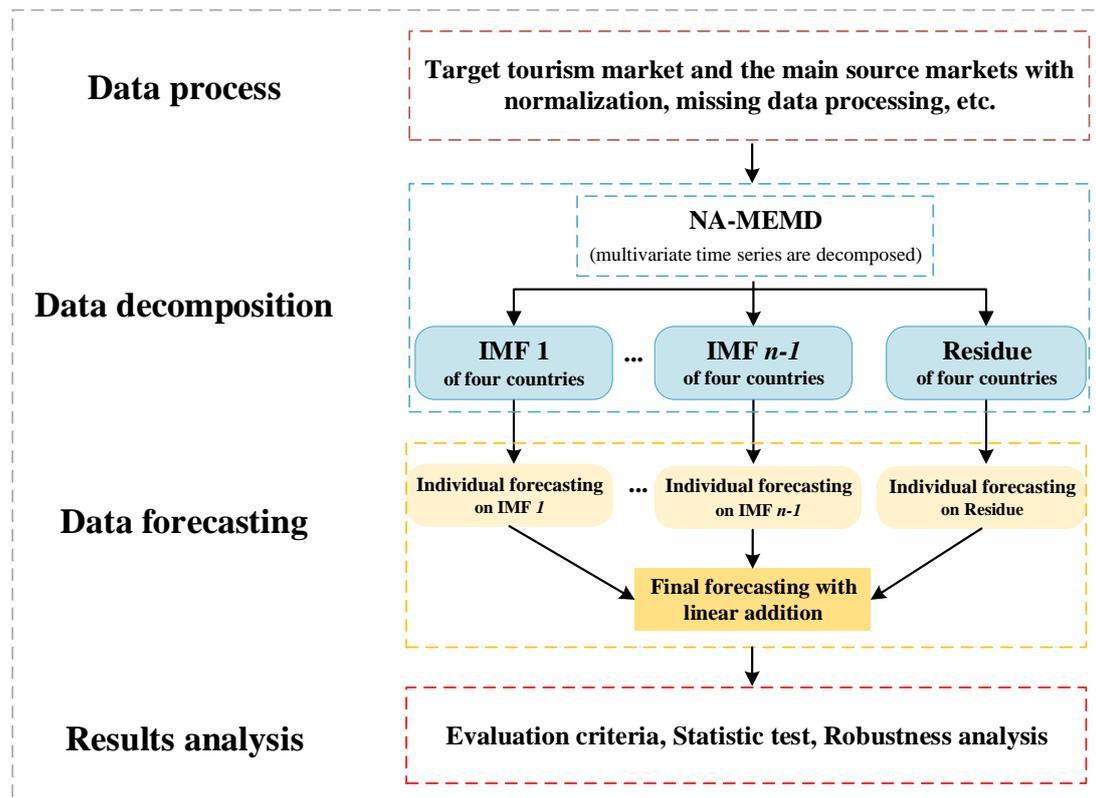

**Fig. 1**. General framework of the proposed forecasting methodology considering main source markets.

(1) **Data process:** $m$-dimension multivariate data $x(t)$, ($t=1,\ldots,T$) including the target tourism market and the corresponding main source markets are determined, and the basic data process with normalization and missing data process for the multivariate time series are accordingly conducted.

(2) **Data decomposition:** By introducing the effective data decomposition technique, i.e., NA-MEMD, every time series of $x(t)$ is consistently and simultaneously decomposed into $n$ intrinsic mode function (IMF) components $c_{j,t}$ ($j=1,2,…,n$) and one residue ($r_t$) and then grouped the components based the timescale.

(3) **Data forecasting:** the individual forecasting of the grouped decomposed components (i.e., $c_{j,t}$ and $r_t$) is conducted with different forecasting techniques (i.e., LR, SARIMA, BPNN, ELM, RVFL and LSTM), and then the final forecasting result is obtained by ensemble each individual forecasting result.

(4) **Result analysis:** the evaluation criteria (i.e., *MAPE*, *RMSE* and $D_{stat}$) in terms of level accuracy and directional accuracy are used for verifying the effectiveness of the proposed approach, and the statistic test is conducted for demonstrating the statistical significance. Moreover, the robustness analysis is used to prove the stability of the decomposition ensemble approach.

## 3.2 Noise-assisted multivariate empirical mode decomposition

The method of NA-MEMD is a multivariate expansion algorithm of empirical mode decomposition (EMD) by using noise-assisted analysis method, which produces the same number of IMFs in all time series data (Zhang et al., 2017b). Considering the mode-alignment (common frequency scales in the same IMF across different time series), NA-MEMD can effectively extract the common factors from interrelated multivariate data at similar timescales, and can adaptively decompose multivariate data and display the amplitude and frequency information of different modal components simultaneously.

Particularly, in terms of data decomposition, the NA-MEMD decomposes multivariate data into several IMF groups, and each IMF group has the same length and components containing the same frequency distribution in the same order of the group (Rehman & Mandic, 2011; Tang et al., 2020). Given $m$-dimension multivariate data $x(t)$, ($t=1,…,T$) where $t$ is the time, NA-MEMD extracts scale-aligned IMFs and residues, by the following processes:

(1) Generating $h$-dimensional uncorrelated Gaussian white noise $n(t)$ of the same length

with x(t), and $n(t) = [n_1(t), n_2(t), \cdots, n_h(t)]^T$.

(2) Adding the n(t) to x(t) and constructing the multivariate signals with (m+h)-dimensional, that is, $z(t) = [z_1(t), z_2(t), \cdots, n_{m+h}(t)]^T$.

(3) Producing K direction vectors $u^k$, (k=1,…,K) based on uniformly distributed points in the (m+h)-dimensional.

(4) Making the projections $p^k(t)$ of multivariate input data z(t) along the direction vector $u^k$ for all k.

(5) Getting the local maxima of the projections $p^k(t)$ and the corresponding the instants $t^k$ at time t in direction k.

(6) Fitting the multivariate envelope curves $e^k(t)$ in a spline form.

(7) Calculating the mean of the envelope curves: $m(t) = \frac{1}{K}\sum_{k=1}^{K} e^k(t)$.

(8) Extracting the detail $d_i(t) = z(t) - m(t)$, if $d_i(t)$ fulfils the criterion for being a multivariable IMFs, then $d_i(t)$ is considered as the i-th IMF component; Computing the residual $r_i(t) = z(t) - d_i(t)$, and repeating steps (4) to (8) for $r_i(t)$, stopping screening when $r_w(t)$ becomes a monotonic function, otherwise, applying the above procedure to $r_i(t)$.

### 3.3 Forecasting techniques

In this section, the forecasting techniques include statistical methods, machine learning and deep learning techniques that are popular in the tourism demand forecasting are introduced as follows (Song et al., 2019):

**Linear Regression (LR)**. LR is one of a basic statistical forecasting model, which has been popularly employed in field of forecasting science (Tang et al., 2020; Shen et al., 2011).

**Seasonal autoregressive integrated moving average (SARIMA)**. SARIMA is one of the popular forecasting methods of econometric model, considering the seasonal effect on the time series (Song & Li, 2008).

**Support Vector Regression (SVR)**. SVR is a typical artificial intelligence tool and has been extensively introduced into the tourism demand forecasting, by mapping the input data into a high-dimension space and minimizing the generalization error (Cortes & Vapnik, 1995).

**Back Propagation Neural Network (BPNN)**. BPNN is one of the most commonly used feedforward neural networks (Tang et al., 2020). BPNN has been applied for modelling the nonlinear data in tourism research (Li et al., 2018).

**Extreme Learning Machine (ELM)**. ELM is an extended version of single-hidden layer feed-forward neural networks without tune the weights and biases iteratively, charactering fast learning speed and generalization performance (Huang et al., 2006). ELM has been employed in the field of tourism demand forecasting (Sun et al., 2019).

**Random Vector Functional Link (RVFL)**. RVFL is similar to ELM using randomly fixed parameters, meanwhile, the RVFL set a link from the input layer to the output layer directly (Pao et al., 1994). RVFL has the ability in dealing with the nonlinear problems of tourism demand forecasting and has been widely used for complex time series forecasting (Tang et al., 2020).

**Long short-term memory (LSTM)**. LSTM network is a promising deep learning technique for modelling nonlinear and nonstationary data, and has also been adopted in tourism demand forecasting (Law et al., 2019). As shown in **Fig. 2**, the LSTM unit is composed of cell, forget gate, input gate and output gate. Accordingly, the cell remembers values over arbitrary time intervals and the three gates regulate the flow of information into and out of the cell, the details are as follows (Law et al., 2019):

(1) Forget gate is used to decide what information will be discarded from the previous cell state

$$f_t = \sigma(W_f \bullet [h_{t-1}, x_t] + b_f), \tag{1}$$

where $h_{t-1}$ represents the output of previous cell state, $x_t$ denotes the input of current cell state, $W_f$ and $b_f$ are the weights and bias, and $\sigma$ refers to sigmoid function.

(2) Input gate will determine how much new information is added to the cell state

$$i_t = \sigma(W_i \bullet [h_{t-1}, x_t] + b_i), \tag{2}$$

$$\tilde{C}_t = \tanh(W_C \bullet [h_{t-1}, x_t] + b_C), \tag{3}$$

$$C_t = f_t \odot C_{t-1} + i_t \odot \tilde{C}_t, \tag{4}$$

where $W_i$ and $b_i$ are the weights and bias of the input gate, $W_C$ and $b_C$ are the weights and bias of the cell state, tanh is the activation function, ⊙ represents point-wise multiplication.

(3) Output gate controls the current information in the cell state to flow into the outputs

$$o_t = \sigma(W_o \bullet [h_{t-1}, x_t] + b_o),  \quad (5)$$

$$h_t = o_t \odot \tanh(C_t),  \quad (6)$$

where $W_C$ and $b_C$ are the weights and bias of the output gate, $O_t$ is used to evaluate which part of cell state to be exported, and $h_t$ calculates the final outputs.

(4) Finally, the forecasted value at $t+1$ is represented by a fully connected layer

$$y_{t+1} = W_{fC} h_t + b_{fC}, \quad (7)$$

where $W_{fc}$ and $b_{fc}$ are the weights and bias of the fully connected layer.

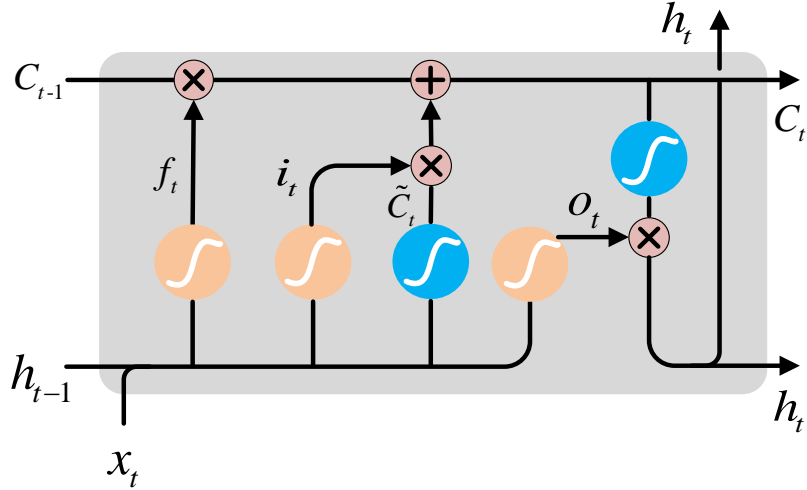

**Fig. 2**. The neural network architecture of LSTM network.

## 4. Experimental design

### 4.1 Data collection

In this study, two countries' inbound tourist volume (i.e., China and Malaysia) were selected as the case study in the proposed approach. The main reason for choosing the above countries as study cases was because they represent countries with rich tourism resources in East and South Asia, respectively, and have reliable tourism data. Notably, China is one of the top ten destinations by international tourist arrivals and Malaysia shares 7.4% of international tourist arrivals in South-East Asia. Therefore, we consider monthly tourist arrivals from top three origin countries (i.e., Japan, South Korea, Russia) to China over the period Jan. 2006 to Feb.

2016 (with 122 samples). Meanwhile, the corresponding monthly inbound tourist volume of three main market source to Malaysia are Singapore, Indonesia and China during Jan. 1999 to Jun. 2019 (with 246 samples). All the eight time series data were collected from Wind Database (http://www.wind.com.cn/). In the empirical study, the times series data were all divided into training dataset (containing the preceding 80% of the total sample) and testing dataset (containing the last 20% of the total sample) in the constructed forecasting models. To verify the robustness of the proposed methods, multistep ahead forecasting at the horizons of 1-3 month(s) were performed.

**4.2 Evaluation criteria**

To measure forecasting performance, the popular three criteria were adopted to evaluate the level and directional forecasting accuracy. As for the level forecasting accuracy, the mean absolute percent error (*MAPE*) and the root mean squared error (*RMSE*) were chose as evaluation criteria (Li & Law, 2019):

$$MAPE = \frac{1}{T}\sum_{t=1}^{T}\left|\frac{y_t - \hat{y}_t}{y_t}\right|, \tag{8}$$

$$RMSE = \sqrt{\frac{1}{T}\sum_{t=1}^{T}(\hat{y}_t - y_t)^2}, \tag{9}$$

where *T* denotes the number of testing dataset, and $\hat{y}_t$ and $y_t$ represent the forecasted and actual tourist arrivals at time *t*, respectively.

As for the directional forecasting accuracy, the directional statistic ($D_{stat}$) was selected for testing the capability to forecast direction of movement (Tang et al., 2020).

$$D_{stat} = \frac{1}{T}\sum_{t=1}^{T}a_t \times 100\%, \tag{10}$$

where $a_t = 1$ if $(\hat{y}_{t+1} - y_t)(y_{t+1} - y_t) \geq 0$, or $a_t = 0$ otherwise.

To verify the superiority of the proposed approach from a statistical perspective, the popular statistic method (i.e., Diebold-Mariano (DM) test) was adopted to test the statistical significance of all models using the mean square prediction error as the loss function, with the null hypothesis that the comparison forecasting models appear a similar forecast accuracy (Diebold & Mariano, 2002). In this study, the DM statistic can be defined as (Sun et al., 2019):

$$S = \frac{\bar{g}}{(\hat{V}_{\bar{g}}/N)^{1/2}}, \qquad (11)$$

where $\bar{g} = 1/N \sum_{t=1}^{N} [(x_t - \hat{x}_{A,t})^2 - (x_t - \hat{x}_{B,t})^2]$, $\hat{V}_{\bar{g}} = \gamma_0 + 2\sum_{l=1}^{\infty} \gamma_l, (\gamma_l = \text{cov}(g_t, g_{t-l}))$, and $\hat{x}_{A,t}$ and $\hat{x}_{B,t}$ denote the forecasts for $x_t$ generated by the proposed approach and the benchmark model B, respectively, at time $t$.

### 4.3 Model specification

As for the model specification in single forecasting models, the SARIMA is constructed according to the Akaike info criterion (AIC) (Sun et al., 2019). In BPNN, a usual three-layer network is built with 7 hidden nodes, and the maximum number of iterations is 10. In ELM and RVFL, the sigmoidal function (i.e., $f(x) = 1/1+e^{-x}$) are respectively selected as the activation function, and the Gaussian distribution are also chosen for producing the input weight and hidden bias randomly (Tang et al., 2020). The number of iterations of ELM and RVFL are both 100. Meanwhile, we set the number of LSTM hidden unit is 64. During the training phase, the train epoch is 300, the solver is 'Adam', the learning rate is 0.01. Due to the small amount of data, we use spatial dropout in the input layer during the training process, we set the dropout rate is 0.5 (Zhao et al., 2017).

## 5. Empirical study

For illustration and verification, the data decomposition results by the NA-MEMD-based method and the comparison of finally forecasting results for two case studies (i.e., China market and Malaysia market) were presented in **Section 5.1** and **Section 5.2**, respectively. **Section 5.3** summarized the major conclusions of the empirical results.

### 5.1 Decomposition results

The decomposition results of monthly time series data of the two case studies by using the NA-MEMD-based method were demonstrated for analyzing the effect of complex impact factors on tourism demand. **Figs. 3** and **4** show the IMF components and the residue, which were all extracted from the two case studies' original data, were sorted from the highest frequency to the lowest frequency, respectively. Moreover, two measures, involving mean period, Pearson correlation, were also calculated for analyzing the dominant mode, as shown in **Tables 2** and **3**.

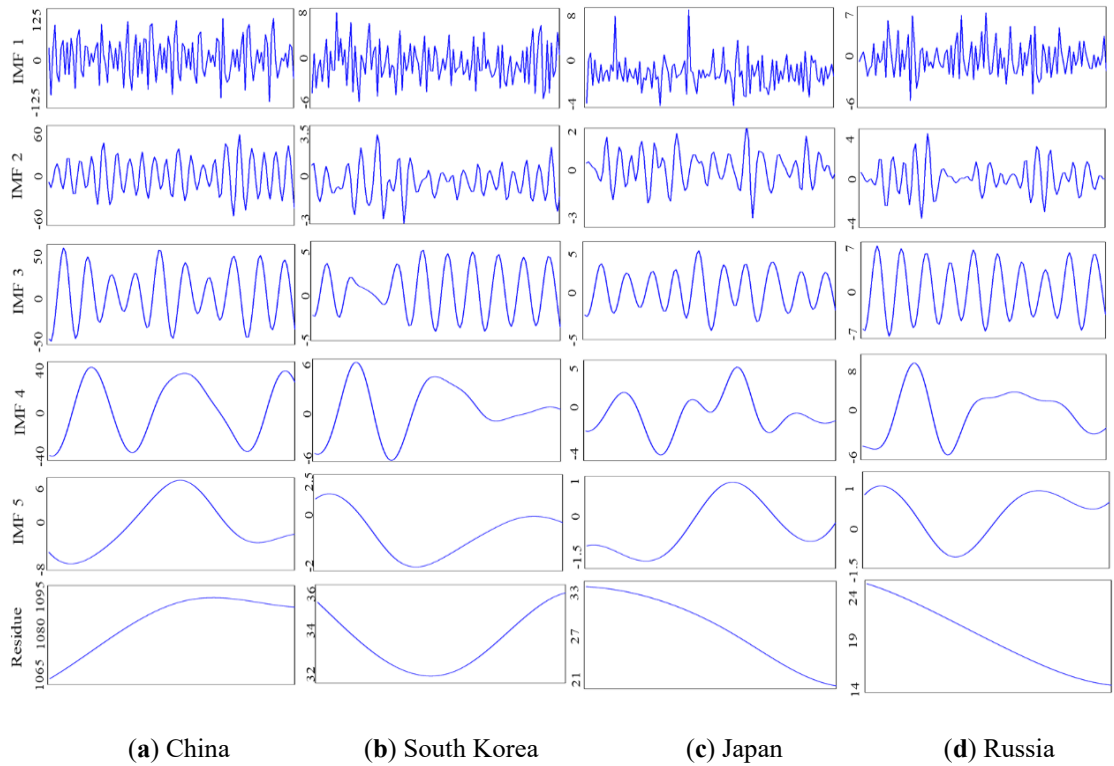

(**a**) China          (**b**) South Korea        (**c**) Japan          (**d**) Russia

**Fig. 3**. Scale-aligned modes of China and three major source countries extracted by NA-MEMD.

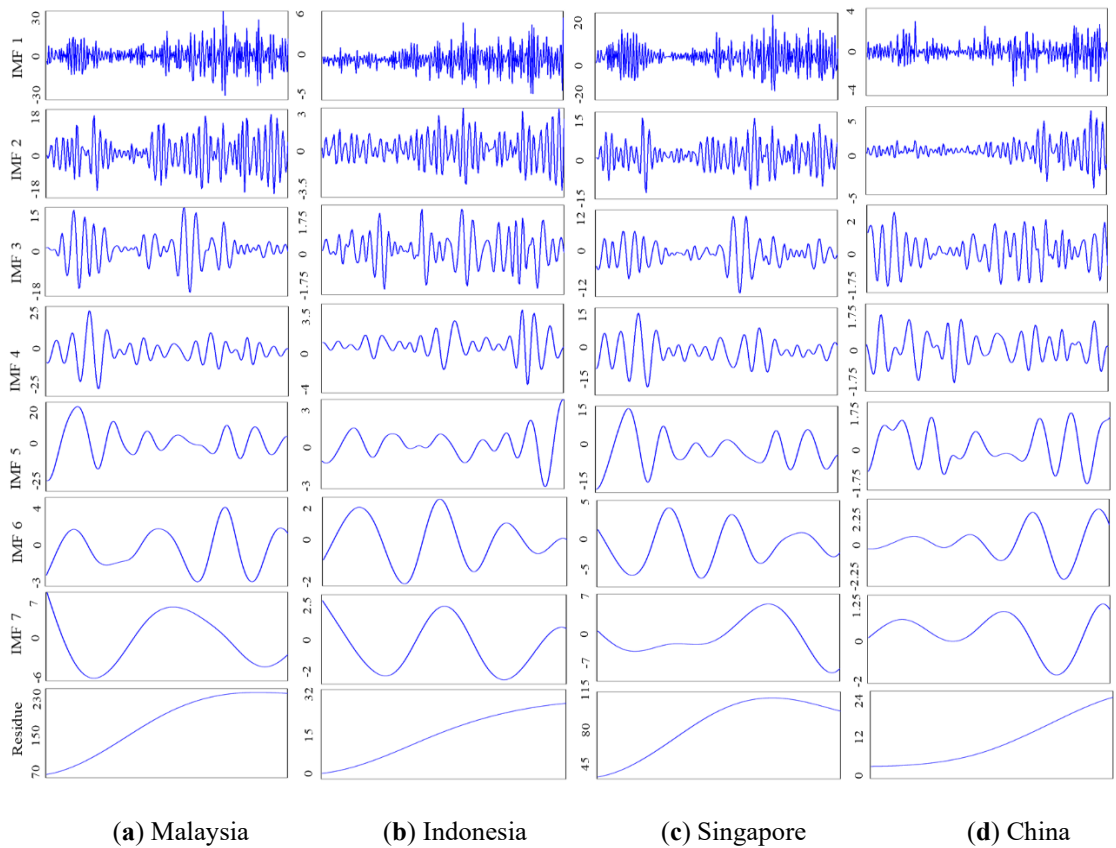

(**a**) Malaysia        (**b**) Indonesia        (**c**) Singapore        (**d**) China

**Fig. 4**. Scale-aligned modes of Malaysia and three major source countries extracted by NA-MEMD.

**Table 2**. Measures of decomposition components in China market.

| Countries | Components | Mean period (Month) | Pearson Correlation |
|---|---|---|---|
| China | IMF 1 | 2.98 | 0.72 |
| | IMF 2 | 6.1 | 0.38 |
| | IMF 3 | 12.2 | 0.35 |
| | IMF 4 | 40.67 | 0.44 |
| | IMF 5 | 122 | 0.26 |
| | Residual | | 0.23 |
| Korea | IMF 1 | 2.98 | 0.53 |
| | IMF 2 | 6.1 | 0.27 |
| | IMF 3 | 12.2 | 0.45 |
| | IMF 4 | 40.67 | 0.57 |
| | IMF 5 | 122 | 0.35 |
| | Residual | | 0.29 |
| Japan | IMF 1 | 2.39 | 0.32 |
| | IMF 2 | 8.71 | 0.2 |
| | IMF 3 | 12.22 | 0.34 |
| | IMF 4 | 61 | 0.47 |
| | IMF 5 | 122 | 0.13 |
| | Residual | | 0.7 |
| Russia | IMF 1 | 2.39 | 0.24 |
| | IMF 2 | 6.1 | 0.25 |
| | IMF 3 | 12.22 | 0.59 |
| | IMF 4 | 40.67 | 0.51 |
| | IMF 5 | 122 | 0.04 |
| | Residual | | 0.4 |

Table 3. Measures of decomposition components in Malaysia market.

| Countries | Components | Mean period (Month) | Pearson Correlation |
|---|---|---|---|
| Malaysia | IMF 1 | 3 | 0.15 |
| | IMF 2 | 6 | 0.13 |
| | IMF 3 | 10.25 | 0.14 |
| | IMF 4 | 17.57 | 0.17 |
| | IMF 5 | 35.14 | 0.08 |
| | IMF 6 | 82 | 0.1 |
| | IMF 7 | 246 | 0.18 |
| | Residual | | 0.95 |
| Indonesia | IMF 1 | 3 | 0.22 |
| | IMF 2 | 6 | 0.14 |
| | IMF 3 | 13.67 | 0.08 |
| | IMF 4 | 14.47 | 0.11 |
| | IMF 5 | 41 | 0.15 |
| | IMF 6 | 82 | 0.02 |
| | IMF 7 | 123 | 0.05 |
| | Residual | | 0.94 |
| Singapore | IMF 1 | 3 | 0.21 |
| | IMF 2 | 6 | 0.2 |
| | IMF 3 | 12.95 | 0.22 |
| | IMF 4 | 17.57 | 0.2 |
| | IMF 5 | 49.2 | 0.14 |
| | IMF 6 | 61.5 | 0.23 |
| | IMF 7 | 246 | 0.45 |
| | Residual | | 0.9 |
| China | IMF 1 | 3 | 0.14 |
| | IMF 2 | 6 | 0.2 |
| | IMF 3 | 10.25 | 0.13 |
| | IMF 4 | 14.47 | 0.11 |
| | IMF 5 | 35.14 | 0.13 |
| | IMF 6 | 61.5 | 0.32 |
| | IMF 7 | 123 | -0.02 |
| | Residual | | 0.93 |

From **Figs. 3** and **4** and **Tables 2** and **3**, one important conclusion based the two kind of measures can be deduced is that the emerging multi-scale analysis, NA-MEMD, effectively captures the common factors that are hidden in the multivariate time series data of the tourist destination and the corresponding major tourism inbound markets, with timescales that are similar within a group but that are different across groups. First, the corresponding decomposed components in the two case studies are all have similar mean period which reveal the common influential factors' impact on tourist arrivals. Second, from the perspective of correlation coefficients, almost all components show a positive correlation with the original tourist arrivals data. Such a satisfactory analysis, therefore, facilitates modelling the relationship between the tourist destination and the corresponding major tourism inbound markets, thereby enhancing the accuracy of the tourist arrivals forecasting.

**5.2 Empirical results**

A comprehensive comparison of the proposed approach in China and Malaysia market with the single forecasting method are demonstrated in this section. For a clear discussion, the compared results between the proposed NA-MEMD-based approach and the corresponding single forecasting model (i.e., LR, SARIMA(X), SVR, BPNN, ELM, RVFL and LSTM) are conducted in case studies in terms of the level and directional forecasting accuracy. Accordingly, the statistical test results are also given.

**5.2.1 China market**

Focusing on the forecasting results, the **Table 4** shows the comparison result with one-, two- and three-step-ahead forecasting from the perspective of *MAPE*, *RMSE* and $D_{stat}$, respectively, and **Table 5** demonstrates the DM test's result for verifying the validity of the proposed approach statistically. From the evaluation criteria and the statistical test, the important conclusion can be clearly deduced that the proposed approach (i.e., NA-MEMD-based model) can be consistently proved to outperform the corresponding single forecasting models in most cases, at the confidence level of 90% based DM test.

Focusing on the *MAPE* and *RMSE* in **Table 4**, the proposed NA-MEMD-based approach outperforms all the corresponding benchmarks in all horizons excepting the RVFL in three-step-ahead forecasting, testifying to its effectiveness for tourism demand forecasting in terms of level forecasting accuracy. Particularly, the proposed approaches, i.e., NA-MEMD-LR, NA-MEMD-SARIMA(X), NA-MEMD-SVR, NA-MEMD-BPNN, NA-MEMD-ELM, NA-MEMD-RVFL and NA-MEMD-LSTM reduce the *MAPE* (and *RMSE*) by approximately 20.4% (22.4%), 42.5% (29.9%), 46.4% (36.9%), 30.2% (33.2%), 31.7% (36.9%), 0.39% (12.0%) and 77.9% (93.3%) on average, respectively, by introducing decomposition ensemble learning approach relative to the single forecasting models. Meanwhile, as for the $D_{stat}$, the proposed approaches with different forecasting models also have a better forecasting performance than the single forecasting models, as shown in **Table 4**. Furthermore, The DM test statistically confirms the superiority of data decomposition methods over single methods in improving forecasting performance with most forecasting techniques under a confidence level of 90% as shown in **Table 5**.

Table 4. Performance comparison of different methods in terms of criteria in China market.

| Horizons | Models | Criteria | Forecasting techniques | | | | | | |
|---|---|---|---|---|---|---|---|---|---|
| | | | LR | SARIMA(X) | SVR | BPNN | ELM | RVFL | LSTM |
| One | Single | MAPE | 0.0313 | 0.0539 | 0.0285 | 0.0280 | 0.0261 | 0.0255 | 0.0316 |
| | | RMSE | 40.13 | 69.63 | 36.21 | 38.29 | 38.30 | 37.53 | 45.48 |
| | | $D_{stat}$ | 0.916 | 0.500 | 0.875 | 0.875 | 0.916 | 0.916 | 0.75 |
| | Proposed | MAPE | **0.0244** | **0.0381** | **0.0194** | **0.0214** | **0.0200** | **0.0247** | **0.0154** |
| | | RMSE | **34.28** | **53.66** | **28.95** | **27.84** | **28.55** | **33.50** | **21.82** |
| | | $D_{stat}$ | **0.916** | **0.708** | **0.958** | **0.958** | **0.958** | **0.916** | **1** |
| Two | Single | MAPE | 0.0305 | 0.0631 | 0.0270 | 0.0306 | 0.0281 | 0.0267 | 0.0320 |
| | | RMSE | 44.28 | 80.23 | 34.18 | 41.16 | 42.01 | 40.45 | 55.15 |
| | | $D_{stat}$ | 0.875 | 0.541 | 0.9166 | 0.875 | 0.916 | 0.916 | 0.875 |
| | Proposed | MAPE | **0.0234** | **0.0375** | **0.0183** | **0.0203** | **0.0199** | **0.0239** | **0.0186** |
| | | RMSE | **32.18** | **53.96** | **25.19** | **27.56** | **28.53** | **31.71** | **25.37** |
| | | $D_{stat}$ | **0.916** | **0.708** | **1** | **0.958** | **0.916** | **0.916** | **0.958** |
| Three | Single | MAPE | 0.0288 | 0.0457 | 0.0246 | 0.0233 | 0.0248 | 0.0241 | 0.0297 |
| | | RMSE | 39.69 | 60.67 | 35.23 | 31.57 | 36.93 | 35.82 | 39.00 |
| | | $D_{stat}$ | 0.875 | 0.625 | 0.833 | 0.916 | 0.875 | 0.875 | 0.875 |
| | Proposed | MAPE | **0.0274** | **0.0385** | **0.0170** | **0.0211** | **0.0200** | **0.0274** | **0.0183** |
| | | RMSE | **34.86** | **54.38** | **23.00** | **27.91** | **28.52** | **35.57** | **25.01** |
| | | $D_{stat}$ | **0.916** | **0.75** | **1** | **0.958** | **0.958** | **0.875** | **0.958** |

Table 5. Results of the DM test between proposed approach and benchmarks in terms of S-statistic (*p*-value) in China market.

| Targets | Benchmark | Forecasting technique | | | | | | |
|---|---|---|---|---|---|---|---|---|
| | | LR | SARIMA | SVR | BPNN | ELM | RVFL | LSTM |
| Proposed model | Original model | -1.52 (0.13) | -3.57 (0.00) | -2.84 (0.00) | -2.70 (0.00) | -2.23 (0.02) | -0.91 (0.36) | -2.41 (0.01) |

#### 5.2.2 Malaysia market

Similarly, the different forecasting results of the proposed approach and the corresponding single model will be compared in terms of the three criteria, as shown in **Table 6**. Meanwhile, the DM test is applied to evaluate forecasting accuracy of different models from a statistical perspective (see **Table 7**). Obviously, in the Malaysia market, the proposed approach is also uniformly superior to the corresponding single model under a confidence level of 90%.

Particularly, as for the level accuracy (i.e., *MAPE* and *RMSE* in **Table 6**), the proposed NA-MEMD-based approach outperforms all the corresponding benchmarks under study in all horizons excepting the SVR in three-step-ahead forecasting, demonstrating the effectiveness of introducing decomposition method to improve the performance of level forecasting. The proposed NA-MEMD-based approach are respectively far smaller the *MAPE* (and *RMSE*) by approximately 37.2% (34.0%), 71.9% (73.6%), 11.7% (14.7%), 17.5% (17.0%), 29.8% (27.8%), 28.2% (27.3%) and 35.0% (39.3%) on average than the single forecasting models. As for directional forecasting accuracy, the $D_{stat}$ results of the proposed appraoches yield much higher approximately 7.5%, 21.1%, 1.7%, 5.7%, 5.7%, 5.7% and 10.5% than their respective single benchmarks. Meanwhile, The DM test also statistically confirms such an improvement made by using the NA-MEMD-based method under a confidence level of 90% in all cases as shown in **Table 7**.

Table 6. Performance comparison of different methods in terms of criteria in Malaysia market.

| Horizons | Models | Criteria | Forecasting techniques | | | | | | |
|---|---|---|---|---|---|---|---|---|---|
| | | | LR | SARIMA(X) | SVR | BPNN | ELM | RVFL | LSTM |
| One | Single | *MAPE* | 0.0496 | 0.0762 | 0.0466 | 0.0429 | 0.0446 | 0.0453 | 0.0456 |
| | | *RMSE* | 13.80 | 21.60 | 12.58 | 12.13 | 12.40 | 12.63 | 14.00 |
| | | $D_{stat}$ | 0.8367 | 0.7551 | 0.7959 | 0.7959 | 0.8571 | 0.8571 | 0.7959 |
| | Proposed | *MAPE* | **0.0274** | **0.0220** | **0.0317** | **0.0288** | **0.0282** | **0.0272** | **0.0231** |
| | | *RMSE* | **7.59** | **5.87** | **8.41** | **8.19** | **7.87** | **7.47** | **6.57** |
| | | $D_{stat}$ | **0.9183** | **0.8571** | **0.8775** | **0.8775** | **0.9183** | **0.9183** | **0.8571** |
| Two | Single | *MAPE* | 0.052529 | 0.0893 | 0.0440 | 0.0430 | 0.0450 | 0.0453 | 0.0514 |
| | | *RMSE* | 13.41512 | 24.68 | 12.16 | 11.93 | 12.09 | 12.13 | 15.77 |
| | | $D_{stat}$ | 0.836735 | 0.6530 | 0.7755 | 0.8367 | 0.8367 | 0.8367 | 0.7551 |
| | Proposed | *MAPE* | **0.0321** | **0.0230** | **0.0403** | **0.0360** | **0.0320** | **0.0325** | **0.0379** |
| | | *RMSE* | **9.25** | **6.05** | **10.8** | **10.48** | **9.14** | **9.32** | **10.66** |
| | | $D_{stat}$ | **0.877551** | **0.8571** | **0.7755** | **0.8775** | **0.8571** | **0.8571** | **0.8775** |
| Three | Single | *MAPE* | 0.050321 | 0.08016 | 0.0421 | 0.0429 | 0.0436 | 0.0439 | 0.0529 |
| | | *RMSE* | 13.51 | 22.58 | 11.87 | 11.98 | 12.15 | 12.24 | 16.07 |
| | | $D_{stat}$ | 0.77551 | 0.7142 | 0.7959 | 0.8367 | 0.7755 | 0.7755 | 0.7755 |
| | Proposed | *MAPE* | **0.0361** | **0.0236** | 0.0450 | **0.0414** | **0.0332** | **0.0368** | **0.0363** |
| | | *RMSE* | **9.98** | **6.21** | 11.97 | **11.23** | **9.43** | **10.08** | **10.58** |
| | | $D_{stat}$ | **0.8367** | **0.8571** | 0.7551 | **0.8571** | **0.8367** | **0.8367** | **0.8367** |

**Table 7.** Results of the DM test between proposed approach and benchmarks in terms of *S*-statistic (*p*-value) in Malaysia market.

| Targets | Benchmark | Forecasting technique | | | | | | |
| --- | --- | --- | --- | --- | --- | --- | --- | --- |
| | | LR | SARIMA | SVR | BPNN | ELM | RVFL | LSTM |
| Proposed model | Original model | -4.577 (0.00) | -8.067 (0.00) | -1.766 (0.07) | -2.174 (0.03) | -4.994 (0.00) | -3.370 (0.00) | -5.293 (0.00) |

### 5.3 Summary

From the empirical results in two cases, four major conclusions can be obtained as follows:

(1) The proposed NA-MEMD-based methodology considering the major source countries significantly outperforms all the corresponding benchmark models for the tourist arrivals in terms of level and directional forecasting accuracy.

(2) Due to the intrinsic non-linear and non-stationary of the tourist arrivals time series, the forecasting results of the AI techniques are overall better than statistical methods (i.e., LR and SARIMA) in tourism demand forecasting.

(3) With the effective decomposition analysis (i.e., NA-MEMD), the novel NA-MEMD-based methodology can be used as an efficient tool for analyzing and forecasting the complex time series, such as tourism demand.

(4) The robustness of the proposed approach has also been proved from the perspective of seven different forecasting techniques, multi-step ahead forecasting and the number of iterations in forecasting techniques.

### 6. Conclusions and managerial implications

By carefully exploring the multi-scale relationship between tourist destination and the major source countries in Asian-pacific region, a novel decomposition methodology (i.e., NA-MEMD-based) is proposed for tourism demand forecasting. In empirical study, through decomposing the tourist arrivals of China and Malaysia market from the corresponding three major source countries, respectively, the similar patterns on tourist arrivals from different influential factors are extracted and identified for ensemble forecasting. The experimental results suggest that the proposed NA-MEMD-based methodology can significantly improve forecasting performance when compared with the benchmarks. As a result, our study makes major contributions to tourism demand forecasting by (1) proposing a novel decomposition ensemble approach for tourist arrivals forecasting, by incorporating the major source countries and exploring their multi-scale relationship with the tourist destination; and (2) introducing the noise-assisted multivariate empirical mode decomposition-based appraoches into tourism demand forecasting; and (3) verifying the effectiveness of the proposed methodology in two cases (i.e., China and Malaysia market) with traditional statistical models, machine learning and deep learning techniques (see **Tables 4** and **6**).

Our research will provide some inspiration in the field of tourism management, especially in terms of tourism demand forecasting. Theoretically, first, it is helpful for researchers and tourism practitioners to understand the main causes of demand fluctuations by exploring the timescale of the complex influencing factors on tourism demand. Furthermore, the integration strategy of linkage decomposition between major source countries and tourist destinations will help to improve the accuracy of tourism demand forecasting. final, accurate tourism demand forecasting will help managers allocate tourism resources efficiently and formulate competitive pricing strategies.

Our proposed methodology could be improved from the following perspectives. In addition to tourist arrivals forecasting, the proposed decomposition ensemble forecasting framework, as an effective analysis and forecasting tool, can be applied to modelling the complex system with non-linear and non-stationary data, such as the crude oil price forecasting and exchange rates forecasting. Moreover, the multi-scale relationship between tourist arrivals and the tourism-related factors can also be explored by the proposed methodology. Third, it would be insightful to investigate study the different time scales of tourism fluctuations caused by various factors (such as political events, war and markets). Fourth, this study only considers the popular data decomposition method to formulate the forecasting framework, i.e., NA-MEMD, however, other more powerful techniques could also be introduced to further enhance the forecasting accuracy. We will investigate these interesting issues in the near future.

## Acknowledgments

This research work was partly supported by the National Natural Science Foundation of China under Grants Nos. 71801213, 71771208, 71642006 and 71988101, and the Research

Grants Council of the Hong Kong Special Administrative Region, China under Grant No. T32-101/15-R.